\documentclass[aps,prd,twocolumn,superscriptaddress,nofootinbib,preprintnumbers,floatfix]{revtex4-2}
\usepackage[normalem]{ulem}
\usepackage{graphicx,hyperref,color,xspace}
\usepackage{subfigure}
\usepackage{amssymb,amsmath}
\usepackage{wasysym}
\usepackage{dcolumn}
\usepackage{bm}
\usepackage{tikz}
\usepackage{pgfplots}
\usepackage{physics}
\usepackage{siunitx}
\usepackage[capitalise]{cleveref}
\pgfplotsset{compat=newest}
\usetikzlibrary{positioning,shapes,arrows,automata}
\usetikzlibrary{decorations.pathmorphing}
\usetikzlibrary{decorations.markings}
\usetikzlibrary{intersections}
\usepackage{circuitikz}
\usepackage{xspace}
\usepackage{lineno}
\usepackage{orcidlink}
\usepackage{makecell}
\usepackage{booktabs}
\usepackage{setspace}
\usepackage{amsthm}
\theoremstyle{definition}

\usepackage{hyperref}
\hypersetup{ 
    pdfnewwindow=true,      
    colorlinks=true,       
    allcolors=[RGB]{31 119 180}
} 

\newcommand{\ie}{\emph{i.e.}\xspace}

\newcommand{\vs}{ \textit{vs.}\xspace}
\newcommand{\eg}{\emph{e.g.}\xspace}

\newcommand{\beq}{\begin{equation}}
\newcommand{\eeq}{\end{equation}}

\DeclareMathOperator{\Det}{Det}

\newcommand{\bea}{\begin{eqnarray}}
\newcommand{\eea}{\end{eqnarray}}

\newcommand{\comma}{\;,}

\newcommand{\V}{\mathcal{V}}

\newcommand{\id}{\mathbf{1}}

\renewcommand{\a}{\alpha}

\newcommand{\n}{\nu}
\newcommand{\m}{\mu}

\newcommand{\e}{\epsilon}
\newcommand{\s}{\sigma}

\newcommand{\oh}{\frac{1}{2}}

\newcommand{\dg}{\dagger}

\newcommand{\ra}{\rightarrow}

\renewcommand{\vec}[1]{\bm #1}
\newcommand{\TwoDVoId}{\texttt{2dVoId}\xspace}
\newcommand{\SUtwo}{SU$(2)$\xspace}
\newcommand{\SUthree}{SU$(3)$\xspace}
\newcommand{\SUN}{SU$(N)$\xspace}

\usepackage[normalem]{ulem}

\newcommand{\ceem}{Center for  Exploration  of  Energy  and  Matter, Indiana  University, Bloomington,  Indiana  47403,  USA}
\newcommand{\indiana}{Department of Physics, Indiana  University, Bloomington,  Indiana  47405,  USA}
\newcommand{\jlab}{Theory Center, Thomas  Jefferson  National  Accelerator  Facility, Newport  News,  Virginia  23606,  USA}
\newcommand{\lbnl}{Nuclear Science Division, Lawrence Berkeley National Laboratory, Berkeley, California 94720, USA}
\newcommand{\ucb}{Department of Physics, University of California, Berkeley, California 94720, USA}
\newcommand{\uned}{Departamento de F\'isica Interdisciplinar, Universidad Nacional de Educaci\'on a Distancia (UNED), Madrid E-28040, Spain}
\newcommand{\sfsu}{Physics and Astronomy Department, San Francisco State University, San Francisco, California 94132, USA}
\newcommand{\wm}{Department of Physics, College of William \& Mary, Williamsburg, Virginia 23187, USA}

\providecommand{\jeff}[1]{}

\begin{document}
\title{First steps towards gauge-independent vortex identification through machine learning}

\preprint{JLAB-THY-26-4764}
\author{Wyatt~A.~Smith\orcidlink{0009-0001-3244-6889}}
\email{wyattsmith@lbl.gov}
\affiliation{\wm}\affiliation{\ucb}\affiliation{\lbnl}
\author{César~Fern\'andez-Ram\'irez\orcidlink{0000-0001-8979-5660}}
\email{cefera@ccia.uned.es}
\affiliation{\uned}
\author{Jeff~Greensite\orcidlink{0000-0003-1720-9436}}
\email{jgreensite@gmail.com}
\affiliation{\sfsu}
\author{Adam~P.~Szczepaniak\orcidlink{0000-0002-4156-5492}}
\affiliation{\jlab}\affiliation{\ceem}\affiliation{\indiana}

\begin{abstract}
As a first step towards machine identification of confining objects in thermalized lattice gauge configurations, we present our \TwoDVoId model for center vortex identification on pure \SUtwo lattices in $D=2$ dimensions.  We create a training set by inserting thin $\mathbb{Z}_2$ vortices at various locations on a zero-action lattice, and then distort those configurations by applying random SU(2) gauge transformations, noise, and by thickening the vortices via cooling.  For moderate vortex visibility, our model is able to reliably identify the location of center vortices. We additionally demonstrate scalability through tiling strategies, which will enable generalization to higher dimensions while reducing training costs.
\end{abstract}
\maketitle

\section{Introduction}\label{Intro}
The center vortex confinement mechanism is one of the leading proposals for how confinement arises in nonabelian gauge theories~\cite{tHooft:1977nqb, Cornwall:1979hz, Nielsen:1979xu,Ambjorn:1980ms}, and lattice simulations have given it substantial support~\cite{DelDebbio:1996lih,DelDebbio:1998luz, Engelhardt:1998wu}. The basic objects are \emph{center vortices}: thin, extended field configurations of codimension 2, carrying a quantized color-magnetic flux valued in the center of the gauge group, \eg $\mathbb{Z}_N$ for \SUN. The picture is that the confining vacuum is densely populated by these vortex sheets. Their effect on a Wilson loop is topological: whenever a vortex worldsheet links the loop---analogous to a flux line piercing a 2D surface bounded by the loop---the loop acquires a multiplicative center element (a phase). If vortex piercings of a minimal surface spanning the loop are sufficiently random and uncorrelated, averaging over these phases produces an area-law decay of the Wilson loop expectation value, \ie, a linear quark--antiquark potential. So confinement, in this picture, comes down to the statistics of how vortices link macroscopic loops. Center vortices have been studied extensively, both on- and off-lattice over several decades~\cite{Kovacs:2000sy,Leinweber:2022ukj,Guvendik:2024umd}. For a broader introduction to the confinement problem and a review of the vortex picture as of 2020, see~\cite{Greensite:2020}.
 
Most of the numerical evidence, however, depends on a single technique: Center vortices are detected on the lattice by taking a configuration, gauge fixing to maximal center gauge, and then center projecting the configuration by replacing each link variable with the nearest element of the $\mathbb{Z}_N$ center of the gauge group~\cite{DelDebbio:1998luz}. The only excitations that survive on the projected $\mathbb{Z}_N$ lattice are center vortices, and the working assumption is that these thin projected vortices lie inside the thicker vortices of the original, unprojected configuration. The procedure has been successful in many respects, but it has always been open to the criticisms that it depends on a particular gauge choice and is therefore subject to the many perils of gauge fixing~\cite{Bornyakov:2000ig,Greensite:2004ur}. Ideally, one would be able to identify vortices directly on a lattice in a random gauge. We know how to insert center vortices into a lattice configuration by hand, completely independent of gauge fixing.  Given that machine learning is very good at pattern recognition~\cite{LeCun:2015pmr}, it is natural to ask whether a neural network, trained on configurations with vortices inserted in known places, could learn to pick out center vortices (or, in principle, other topological objects) in configurations generated by ordinary Monte Carlo. Related work has applied machine learning to lattice gauge configurations, from phase identification and order parameters~\cite{Carrasquilla:2017,Wetzel:2017} to symmetry-aware neural architectures~\cite{Favoni:2020reg,Nagai:2021bhh,Nagai:2025rok,Tomiya:2025quf}. Applying similar methodologies to center vortex detection would allow one to compute the vortex contribution to the string tension of Wilson loops without ever fixing a gauge. Agreement between that contribution and the known string tension of the theory would eliminate the standard objection to center projection. In this paper we take the first step in that direction. We demonstrate vortex identification on a $D=2$ lattice in \SUtwo, using convolutional neural networks. In what follows, \Cref{sec:confinement_from_cv} gives a brief derivation of the vortex confinement mechanism. \Cref{sec:ML} describes our simulation procedure, the architecture of our 2D Vortex Identification model \TwoDVoId and how it can be scaled to larger lattices. \Cref{sec:results} provides the results. \Cref{sec:conclusions} contains our outlook.

\section{Confinement from center vortices}\label{sec:confinement_from_cv}
In a pure \SUN gauge theory, the asymptotic string tension of Wilson loops in a given representation of the gauge group depends only on the $N$-ality of the group representation, rather than, say, the Casimir of the group representation, which is important at intermediate distances; see~\eg~\cite{Bali:2000gf,Greensite:2020}.  While this fact can be explained in a ``particle'' picture in terms of the screening of loops by gluons, we would also like an explanation in terms of the field configurations that dominate the path integral at large scales.  As far as we are aware, center vortices are the only field configurations whose quantum fluctuations can generate string tensions with the required $N$-ality property, while also giving a natural account of approximate Casimir scaling at intermediate distances through their finite thickness~\cite{Faber:1997rp}.

\begin{figure}
\includegraphics[scale=0.27]{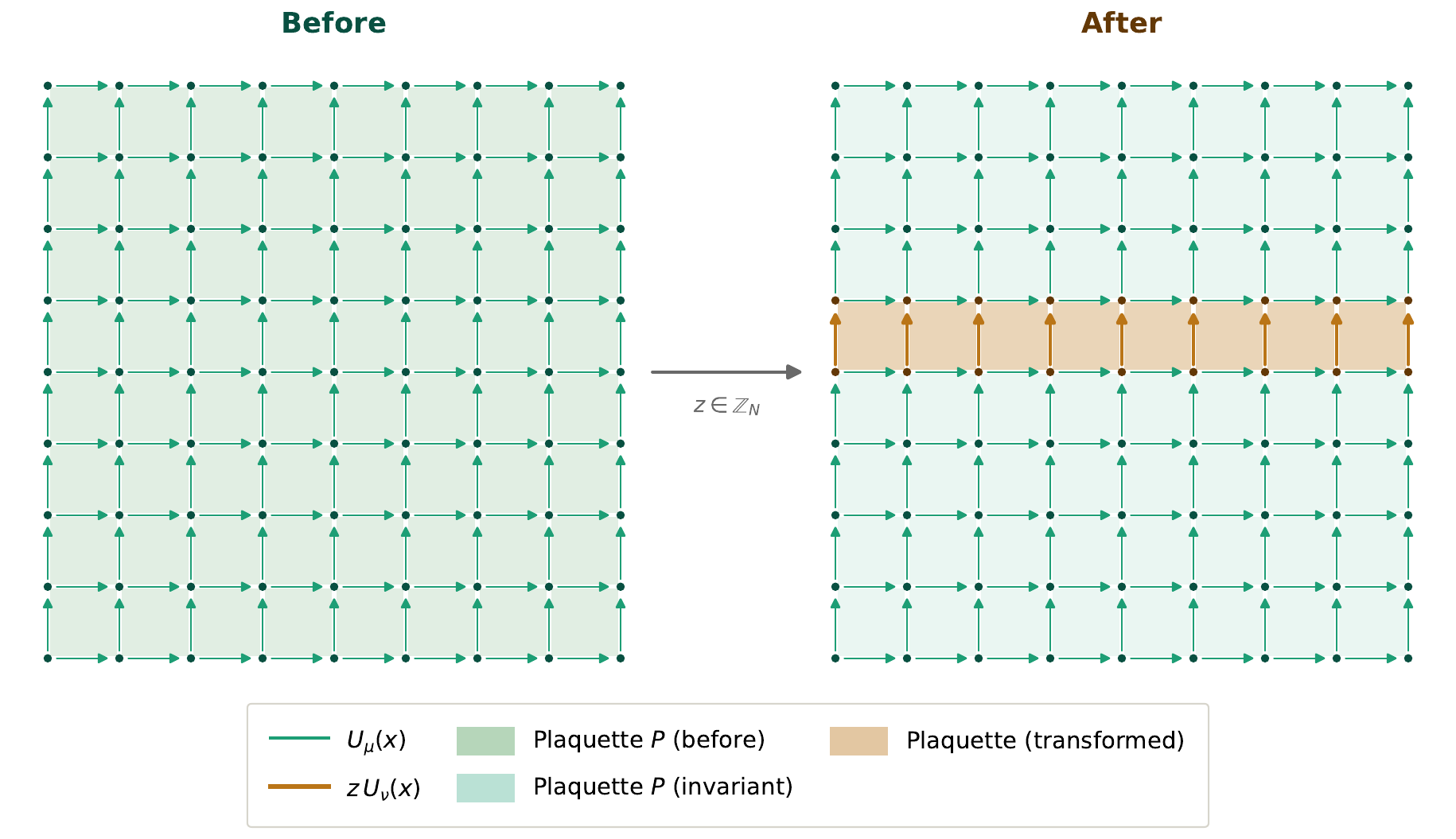}
\caption{A global center transformation.  Each of the indicated links is multiplied by the same center element $z$; the lattice action is unchanged.}
\label{global}
\end{figure}

\begin{figure}
\includegraphics[scale=0.29]{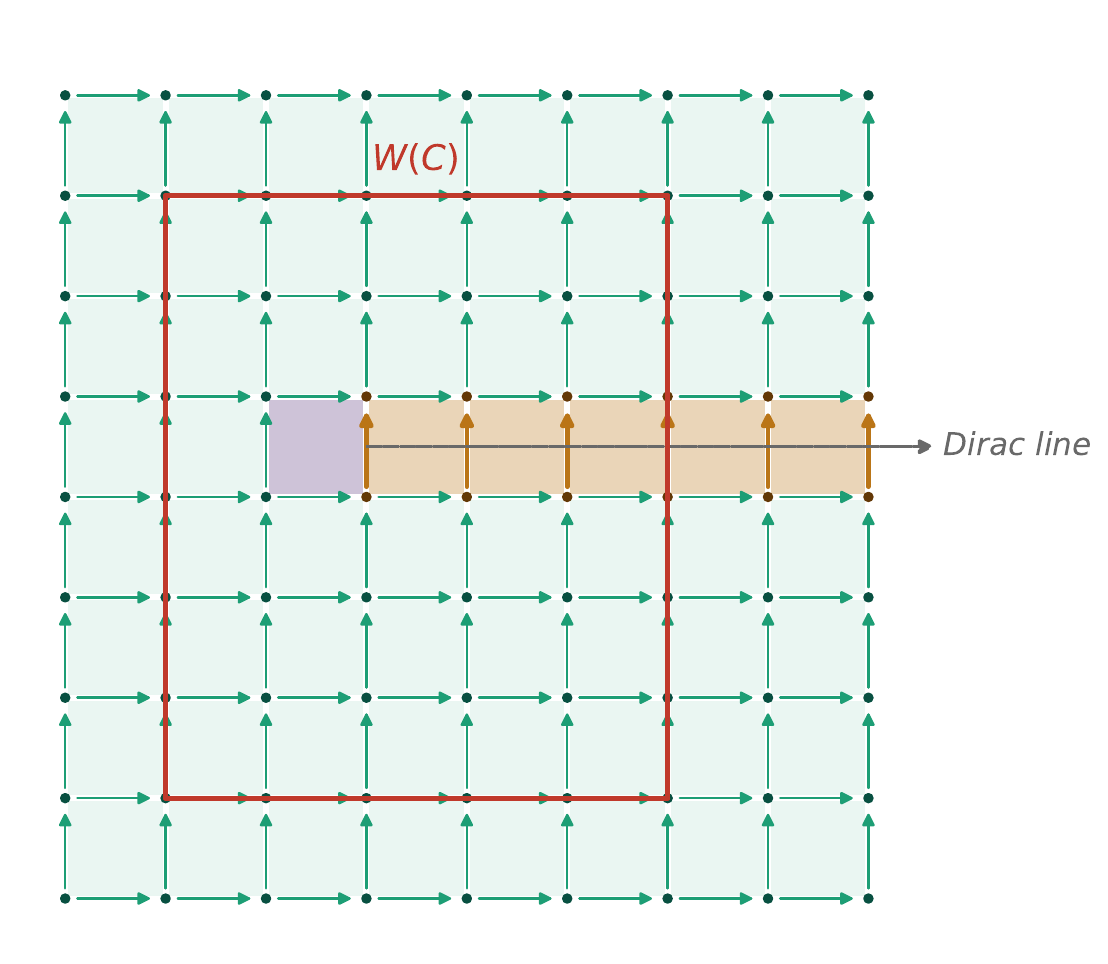}
\caption{Creation of a thin center vortex. The action is changed at the purple-shaded plaquette, and a Wilson loop holonomy, linked to the vortex, is multiplied by the center element $z$.}
\label{cv3}
\end{figure}

It is simplest to introduce center vortices by explaining how they are created in a lattice configuration. All pure \SUN gauge theories are invariant under a one-form symmetry that used to be called ``center symmetry''~\cite{Gaiotto:2014kfa}. Let us begin with a pure \SUN gauge theory in $D=2$ dimensions. \Cref{global} shows an example of center symmetry transformations, in which all time-like links on a certain time slice are multiplied by an element $z \in \mathbb{Z}_N$ of the center of the gauge group. This kind of transformation will multiply Polyakov lines by $z$, but Wilson loops are not affected, since for every link in the loop multiplied by $z$ there is another link multiplied by $z^{-1}$. Consider then the incomplete center transformation shown in \cref{cv3}, where the multiplication by $z$ ends at the shaded plaquette.  Only one link in the plaquette is multiplied by $z$, and so this operation changes the value of the plaquette by a factor $z$.  But this is also true of every Wilson loop that encloses that shaded plaquette, which we now refer to as a thin vortex.  The vortex can be thickened and distorted in various ways, which we will discuss below, without losing the property of multiplying Wilson loops that enclose the vortex by a center element.  As we move to higher dimensions, the concept of ``enclosed'' is generalized to linking.  In two, three, and four dimensions, center vortices are point-like, line-like, or surface-like objects, respectively, and a Wilson loop may link topologically to objects of this kind.  

Let the gauge group be \SUtwo for simplicity. Consider a plane of area $L^2$ which is pierced, at random locations, by $N_v$ center vortices, and consider a Wilson loop $C$ of area $A$ lying in that plane.  Then the probability that $n_c$ of these $N_v$ vortices will lie inside the area $A$ is
\beq
P_{N_v}(n_c) = {N_v \choose n_c}\left( A\over L^2 \right)^{n_c}\left(1 - {A \over L^2} \right)^{N_v-n_c}.
\eeq
Each vortex piercing the Wilson loop contributes a factor of $-1$, so the contribution of the vortex to the Wilson loop is
\beq
 W(C) = \sum_{n_c=0}^{N_v} (-1)^{n_c} P_{N_v}(n_c) = \left(1- {2A \over L^2} \right)^{N_v}.
\eeq
Now, keeping the vortex density $\rho=N_v/L^2$ fixed and taking the limit $N_v, L \ra \infty$, we arrive at the Wilson-loop area-law falloff
\beq
W(C) = \lim_{N_v \ra \infty} \left(1 - {2\rho A \over N_v} \right)^{N_v} = e^{-2 \rho A}.
\eeq
That is the vortex confinement mechanism in three equations~\cite{Engelhardt:1998wu}.  We believe it is the simplest known. The crucial assumption is that vortex piercings in the plane are random and uncorrelated, and this implies, in $D>2$ dimensions, that vortices percolate throughout the spacetime volume~\cite{Engelhardt:1999fd}.

Inserting center vortices into a given lattice configuration is straightforward; identifying the vortices already present in a \mbox{Monte Carlo--generated} configuration is a separate problem. The standard technique, mentioned above, is to fix to the maximal center gauge and project each group-valued link variable to the nearest element of $\mathbb{Z}_N$~\cite{DelDebbio:1998luz}. On the resulting $\mathbb{Z}_N$ lattice, every plaquette whose value differs from unity belongs to a vortex, and the expectation is that these thin projected vortices trace out the cores of the thicker vortices in the underlying gauge field.

For a planar loop with known vortex locations, the vortex contribution to an \SUN loop $C$ in the fundamental representation, for a given lattice configuration, $U_\m(x)$ is obtained by multiplying the $z\in \mathbb{Z}_N$ elements of each vortex that pierces the plane of the loop, \ie ${\cal Z}_C(U)=\prod_i z_i$, where the product runs over the vortices piercing the plane of the loop. Averaging the product ${\cal Z}_C(U)$ over all lattice configurations $U_\m(x)$ generated in a lattice Monte Carlo simulation, and over all loops of the same shape as loop $C$ on the lattice, provides the expectation value of the loop, and hence the vortex contribution to the string tension.

\section{Machine Learning}\label{sec:ML}
Our ultimate goal is to find an alternate method of vortex identification with machine learning, but in order to do that we must set up the learning process by defining our model and supplying a training data set. It should be emphasized from the start that we are not aiming to identify center vortices generated by simulation of \SUtwo lattice gauge theory in two dimensions, because in two dimensions such objects would be uncommon. Instead, the confinement mechanism in two dimensions is, owing to the absence of a Bianchi identity, the lack of plaquette correlation which leads to Casimir scaling rather than $N$-ality dependence~\cite{Greensite:2020}. The idea is instead to train a model to recognize center vortices that are inserted by hand at known positions on a noisy background with \mbox{\SUtwo-valued} link variables.  This is the recognition problem that we expect to encounter in \SUtwo lattice gauge theory in three and four dimensions, when center vortices are indeed frequently generated by a Monte Carlo simulation of the lattice gauge theory.

\subsection{Training and validation data generation}
We construct training configurations containing $N_{\rm flip}$ pairs of center vortices at known positions on 2D lattices. Starting from a ``cold'' lattice with $U_\m(x) = +\id$ on every link, each pair is created by flipping a string of $d+1$ consecutive links of orientation $\m$ from $+\id$ to $-\id$, where $d$ is the desired separation between the two vortices along the direction perpendicular to $\m$.  After all flips, the only plaquettes with value $-\id$ are those at the endpoints of the strings; these are the vortices. The configuration contains exactly $2N_{\rm flip}$ vortices at known positions, with density $2N_{\rm flip}/L^2$. The pair separation $d$ ranges from 1 (adjacent vortices) to $L/2$ (maximally separated on a periodic lattice of size $L$), and for each pair, the separation and string orientation $\m \in \{0,1\}$ are chosen independently. The starting positions are independently randomized.  We train separate models for $N_{\rm flip} \in \{1, 2, 4, 8, 12\}$, corresponding to lattices with $2$, $4$, $8$, $16$, and $24$ vortices on a $24 \times 24$ periodic lattice.

\begin{figure*}[t]
\centering
\includegraphics[width=\textwidth]{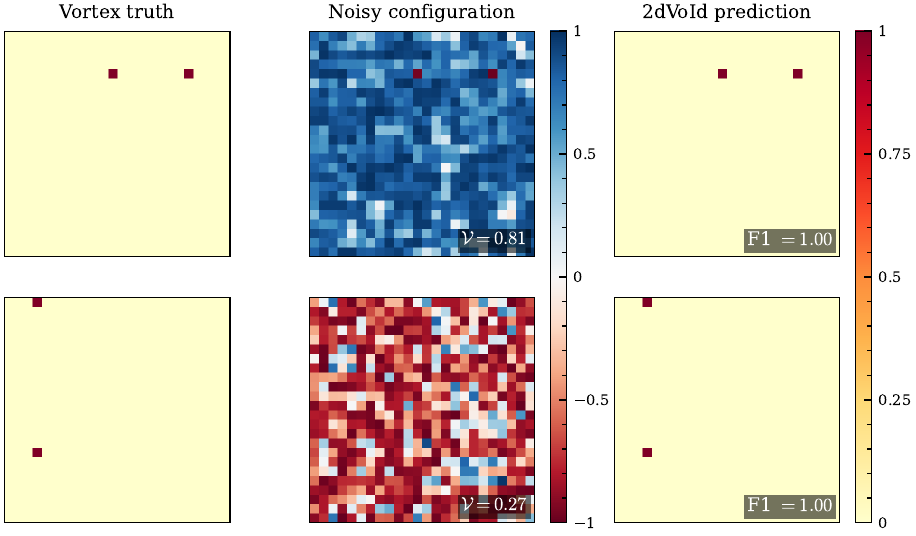}
\caption{Vortex detection at high and low visibility, with $N_{\rm flip} = 1$ (two vortices on a $24 \times 24$ lattice).  Each row shows the ground-truth vortex mask (left), the cooled plaquette trace $\oh\tr(P)$ (center), and the \TwoDVoId probability map $P(\mathrm{vortex})$ (right).  Visibility $\V$ and vortex F1 score are annotated on the center and right panels.}
\label{fig:showcase}
\end{figure*}

After constructing a vortex configuration, we distort it in three ways.  Both of the first two steps use random \SUtwo matrices of the form $\exp(i\,\vec\e \cdot \vec\s/2)$, where $\hat\e$ is a random direction in 3D space, $\vec\s$ are the Pauli matrices, and $|\vec\e|$ is drawn from a half-normal distribution whose width we call the \textit{scale}.

First, we apply a random gauge transformation,
\begin{align}
U_\m(x) \to g(x)\, U_\m(x)\, g^\dg(x+\hat\m) \comma
\end{align}
with one random \SUtwo matrix $g(x)$ per lattice site, at scale $\s_{\rm gauge}$.  Since plaquettes transform as $P_{\m\n}(x) \to g(x) P_{\m\n}(x) g^\dg(x)$, the center element ($\pm\id$) of each plaquette is preserved exactly.  Second, we apply link noise,
\begin{align}
U_\m(x) \to U_\m(x)\, R_\m(x) \comma
\end{align}
with another independent \SUtwo random matrix $R_\m(x)$ per link, at scale $\s_{\rm noise}$.  Unlike the gauge transformation, this noise can distort plaquette values and obscure vortex locations.  The network is trained to predict the ``ground-truth'', \ie the binary label marking which plaquettes are pierced by a vortex in the original configuration, before any gauge transformations or noise were applied.

Finally, we apply iterative cooling to smooth the link variables and thicken the inserted vortices~\cite{Trewartha:2015ida}, \ie,
\beq \label{eq:cooling}
U_\m(x) \;\rightarrow\; (1-\a)\, U_\m(x) \;+\; \frac{\a}{2}\, S_\m(x) \comma
\eeq
followed by rescaling $U \to U / \sqrt{|\Det U|}$ to restore unitarity.  Here $S_\m(x)$ is the sum of staples, and $\a = 0.2$ is the relaxation parameter. We generally performed three steps of cooling. Cooling is intended to make the inserted configurations resemble a two-dimensional spatial slice of the thick center vortices generated by Yang-Mills dynamics in higher dimensions.

By varying the noise scales and the number of cooling steps, each training configuration ends up at a different effective ``difficulty'' of the identification problem.  Rather than characterize this difficulty by the noise parameters themselves, we define a single scalar, the \textit{vortex visibility} $\V$, that captures how distinguishable the vortex signal is in the processed configuration from its plaquettes.  Given the plaquette traces at every lattice site, let $\m_V, \s_V$ denote the mean and standard deviation of these traces at vortex sites, and $\m_N, \s_N$ the corresponding quantities at non-vortex sites.  The effect-size separation is
\beq \label{eq:effectsize}
\Delta = \frac{\m_N - \m_V}{\sqrt{\s_V^2 + \s_N^2}} \comma
\eeq
and we rescale to the unit interval via
\beq \label{eq:visibility}
\V = \frac{\Delta}{\Delta+ a} \comma
\eeq
where $a=2$ was chosen empirically to match our natural intuition about low- and high-vortex visibility. Configurations with $\V$ near 1 have vortex signals well separated from the background; those with $\V$ near 0 have vortex signals buried in noise.

To ensure the trained models generalize across a range of visibilities, we produce training data spanning a grid of scales $(\s_{\rm gauge}, \s_{\rm noise}) \in [0, 0.3] \times [0, 0.3]$ with step size $0.1$, giving $16$ combinations.  For each combination we generate $400$ training lattice configurations and 40 separate validation configurations that are withheld from training and used only to assess generalization.

To give a concrete sense of what the network faces, \cref{fig:showcase} shows two examples at opposite ends of the visibility scale, both with two vortices on a $24 \times 24$ lattice.  In the top row ($\V = 0.81$), the vortex plaquettes are plainly visible as red spots in the cooled plaquette traces.  In the bottom row ($\V = 0.27$), noise has washed out most of the contrast; the input looks nearly uniform to the eye.  We anticipate that our vortex identification model, \TwoDVoId, still correctly and exactly identifies the locations of the vortices, as shown in the right column, which displays our predicted probability $P(\mathrm{vortex})$ at each plaquette site; warmer colors correspond to higher predicted probability on a scale from 0 to 1. The details and full results of \TwoDVoId are provided in the next sections. We quantify detection quality with the vortex F1 score, the harmonic mean of the precision (fraction of flagged plaquettes that are genuine vortices) and the recall (fraction of genuine vortices that are flagged),
\beq \label{eq:f1}
\mathrm{F1} = \frac{2 N_\text{TP}}{2 N_\text{TP} + N_\text{FP} + N_\text{FN}} \comma
\eeq
where $N_\text{TP}$ is the number of correctly flagged vortex sites, $N_\text{FP}$ the number of non-vortex sites incorrectly flagged as vortices, and $N_\text{FN}$ the number of vortex sites the network missed.  F1 $= 1$ corresponds to perfect detection; either a steady stream of false alarms or a steady stream of missed vortices drives it to zero.

\begin{figure*}[t]
\centering
\includegraphics[width=0.9\textwidth]{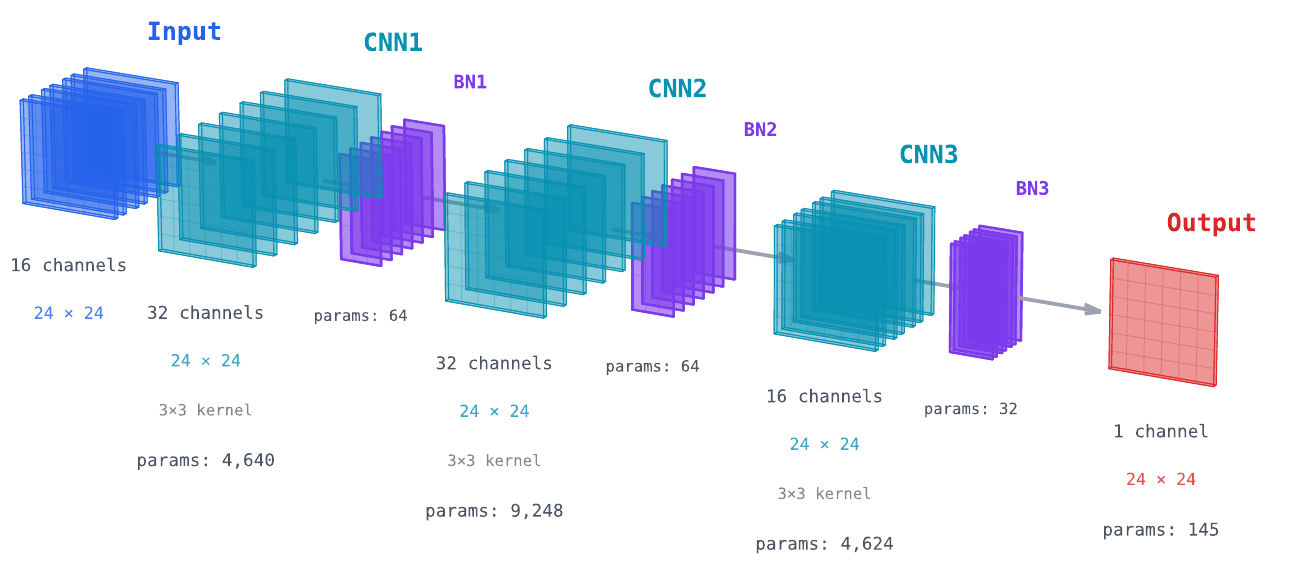}
\caption{\TwoDVoId model architecture based on three hidden layers of convolutional neural networks: \texttt{CNN1}, \texttt{CNN2} and \texttt{CNN3}. After each convolution layer we place a \texttt{ReLU} activation function and we rescale to unit variance and zero mean using \texttt{BatchNorm} (BN). The final output layer applies a sigmoid activation to produce a probability map. All convolutions use $3\times 3$ kernels with circular padding.}
\label{fig:convarch}
\end{figure*}


\subsection{\TwoDVoId architecture}
In this section we present the full architecture of \TwoDVoId, our 2D Vortex Identification model, depicted in~\cref{fig:convarch}. \TwoDVoId employs convolutional neural networks~\cite{LeCun:1998,Goodfellow-et-al-2016} that respect the periodic boundary conditions of the lattice.  The input to \TwoDVoId is the raw lattice configuration $\Lambda = \{ U_\m(x) \}$. Each link $U_\mu(x) \in \mathrm{SU}(2)$ is a $2 \times 2$ complex matrix, which we decompose into its real and imaginary parts to obtain $8$ real numbers per link. In two dimensions there are two links per site (one for each direction $\mu = 0, 1$), giving $16$ real input channels per lattice site $x$.

These input channels form the initial feature map $h^{(c)}(x)$, with channel index $c = 1, \dots, 16$. The network then applies a sequence of convolutional layers. A given layer maps an input feature map with $C_{\rm in}$ channels to an output feature map with $C_{\rm out}$ channels, and is parameterized by a kernel tensor $K$ of shape $(C_{\rm out}, C_{\rm in}, k, k)$, where $k$ is the (odd) spatial extent of the kernel. At each lattice site $x$, the layer computes
\beq
h^{(c)}_{\rm out}(x) = f\!\Bigl( \sum_{c'=1}^{C_{\rm in}} \sum_{i,j = -(k-1)/2}^{(k-1)/2} K^{(c,c')}_{ij}\, h^{(c')}_{\rm in}(x + i\hat 0 + j\hat 1) + b^{(c)} \Bigr) \comma
\eeq
for $c = 1, \dots, C_{\rm out}$, where $h^{(c')}_{\rm in}$ and $h^{(c)}_{\rm out}$ are the input and output feature maps, $\hat 0$ and $\hat 1$ are unit vectors along the two lattice directions, $b^{(c)}$ is a learned bias for output channel $c$, and $f : \mathbb{R} \to \mathbb{R}$ is a nonlinear activation function applied componentwise. The kernel weights $K^{(c,c')}_{ij}$ and biases $b^{(c)}$ are shared across all lattice sites $x$, which enforces translational invariance of the layer. When $x$ lies near a boundary of the lattice, the sum above requires values of $h^{(c')}_{\rm in}$ at sites outside the lattice; the prescription for assigning these values is called \textit{padding}. We use \textit{circular padding}, which wraps the lattice periodically so that a site beyond one edge is identified with the corresponding site on the opposite edge.  This implements the usual periodic boundary conditions exactly.  With kernel size $k=3$, each output site aggregates information from a $3\times 3$ neighborhood of input sites, encompassing the four links comprising a plaquette plus their nearest neighbors.  The kernel weights encode learned linear combinations of the input channels within this neighborhood; different output channels capture different correlation patterns.  Stacking convolutional layers expands the effective correlation length, allowing deeper networks to encode longer-range link correlations. 

Our \TwoDVoId consists of three hidden convolutional layers with $3\times 3$ kernels, as depicted in~\cref{fig:convarch}. The layers are sequential, meaning the number of input channels for each layer is equal to the number of channels in the previous layer.  The intermediate layers use \texttt{ReLU} activations, $f(x) = \max(0,x)$, which introduce nonlinearity.  The final output layer applies a sigmoid activation, $\s(x) = 1/(1+e^{-x})$, producing a smooth probability map $P(x) \in [0,1]$ at each plaquette site indicating the likelihood that the corresponding plaquette is pierced by a vortex.  All code was written using \texttt{PyTorch}, making use of the \texttt{Conv2d} method~\cite{NEURIPS2019_9015}.

We minimize the binary cross-entropy (BCE) loss~\cite{Goodfellow-et-al-2016}
\begin{align} \mathcal{L}_\text{BCE}=-\sum_x \Big( y(x)\ln P(x) + \left[1-y(x)\right]\ln\left[1-P(x)\right]\Big), \end{align} which measures how well the predicted probabilities $P(x)$ match the ground-truth labels $y(x) \in \{0,1\}$.  Optimization uses the \texttt{Adam} algorithm~\cite{Kingma2014AdamAM}, with initial learning rate $10^{-3}$; the learning rate is reduced by a factor of $0.7$ whenever the loss on a held-out validation set (\ie the $40$ configurations per noise combination not used during training) fails to improve for $5$ consecutive epochs.  One epoch is a single pass through the full training set; training proceeds for $30$ epochs with batches of $64$ configurations, and we retain the model weights from the epoch with the lowest validation loss.

\begin{figure}[t]
\centering
\includegraphics[width=\columnwidth]{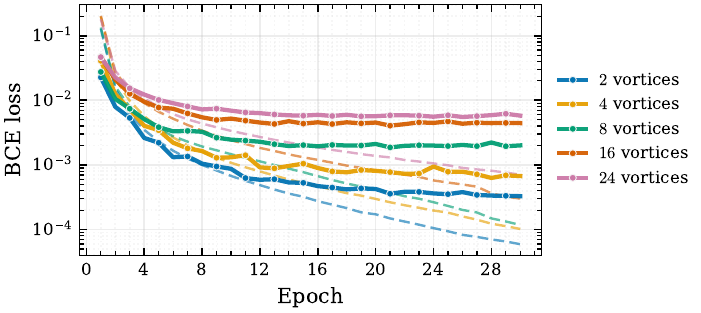}
\caption{BCE loss \vs training epoch for all five vortex densities ($2\dots 24$ vortices). Solid curves show validation loss; dashed curves show training loss.}
\label{fig:training}
\end{figure}

\Cref{fig:training} shows the training and validation loss curves for all $5$ density models. All converge well within $30$ epochs; denser configurations yield higher final loss, which is expected since more vortices on a fixed lattice amounts to a harder classification task. The validation loss is small in every case, which indicates that the models generalize well, though at later epochs the model does begin to overfit the training data as evidenced by the split between the training and validation curves.

\subsection{Tiling to larger lattices}\label{sec:tiling}
Once trained, the same \TwoDVoId model can in principle be applied to lattices of arbitrary size and dimensions. In order to reduce computational costs of vortex detection on larger lattices, we can exploit the translational invariance which we already built into the model itself. If we train on a small lattice, where we can afford to generate large and diverse training sets, and then deploy on larger lattices by ``copying'' the model and tiling the plane (or planes in higher dimensions), we decouple the training cost from the inference volume. Training then scales with the patch size, while inference scales linearly with the lattice volume, as in fully convolutional dense-prediction models~\cite{Long:2015}. This is drastically cheaper than retraining a separate model for every desired lattice size.

The \TwoDVoId model as described above has a receptive field of $9 \times 9$ lattice sites, meaning the classification of each lattice site as containing a vortex or not depends on lattice sites up to $4$ lattice units away in each direction. Since circular padding wraps information around the torus, a model trained on a small enough lattice with circular padding effectively ``sees'' the entire lattice at every site, an advantage that would not carry over to training the model for larger lattices.  To test whether a small model can generalize, we train a second network on $12 \times 12$ lattices using \textit{zero padding} instead of circular padding, so that the network treats patch boundaries as unknown rather than wrapping around to the other side of the lattice. 

At inference time, we tile this $12 \times 12$ model onto a larger lattice by dividing it into patches. On our \mbox{$24 \times 24$} test lattice, we extract four non-overlapping $12 \times 12$ quadrants.  Before feeding a quadrant to the network, we pad it with a halo of 5 border sites on every side, drawn from the neighboring (periodic) lattice region; this gives the network local context at every interior site, analogous to ghost cells in finite-difference methods.  The padded input is therefore $22$ sites on a side, but the model itself was trained only on $12 \times 12$ lattices. The extra border simply attempts to mitigate the effect of zero-padded edges corrupting predictions near the patch boundary.  After the network processes the padded input, we crop away the halo and retain only the central $12 \times 12$ of the output.  The four crops are stitched together to provide the full $24 \times 24$ prediction.

Because the retained sub-lattices tile the full lattice exactly, there is no overlap in the output predictions. The halo regions overlap with neighboring quadrants only when they are used as input to the model, and are discarded after inference.  On a $48 \times 48$ lattice the same $12 \times 12$ patches would tile into 16 non-overlapping quadrants, and on the lattice sizes relevant to $D=3$ and $D=4$ simulations the overhead would be negligible.

\section{Results}\label{sec:results}
\begin{figure}[t]
\centering
\includegraphics[width=\columnwidth]{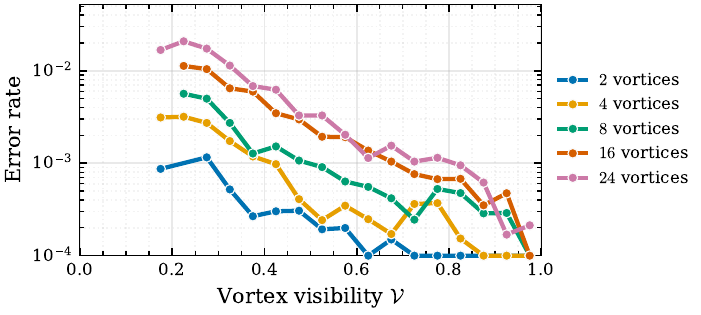}
\caption{Per-plaquette error rate defined as ($1 - \text{accuracy}$) across all training configurations as a function of vortex visibility $\V$.}
\label{fig:acc-vs-vis}
\end{figure}

We now evaluate the trained models across the full range of visibilities and vortex densities; all results in this section are for a $24 \times 24$ lattice. \Cref{fig:acc-vs-vis} shows the per-plaquette error rate as a function of vortex visibility for all $5$ density models. Test configurations are drawn from the same noise and cooling grid used in training, with fresh random seeds, and each point averages over configurations within a narrow visibility bin. The visibility metric captures what it was designed to: curves for all densities collapse onto a broadly common trend, confirming that $\V$ encodes the dominant factor controlling detection difficulty. Error drops by orders of magnitude as $\V$ increases; at high visibility ($\V > 0.6$), all densities achieve sub-1\% error. At low visibility ($\V < 0.3$), however, the curves fan out: the model still performs well with only two vortices, but struggles on 24-vortex configurations where overlapping signals make the task harder.

\begin{figure}[t]
\centering
\includegraphics[width=\columnwidth]{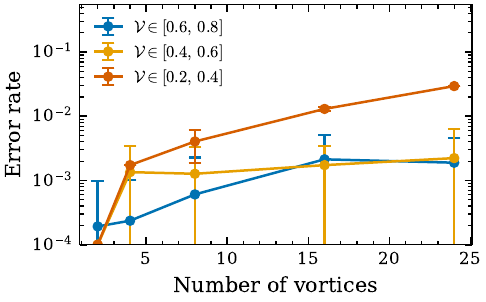}
\caption{Error rate versus number of vortices at three fixed visibility bands.  Even at equal visibility, denser configurations are harder to classify.}
\label{fig:acc-vs-density}
\end{figure}

\Cref{fig:acc-vs-density} shows the overall error rate as a function of vortex density at fixed visibility. We bin test configurations into three visibility bands and plot error rate against the number of vortices. Predictably, denser vortex configurations are more difficult to reconstruct at every visibility level.

\begin{figure*}[p]
\centering
\begin{tabular}{ccc}
\includegraphics[height=0.7\textheight,width=0.32\textwidth]{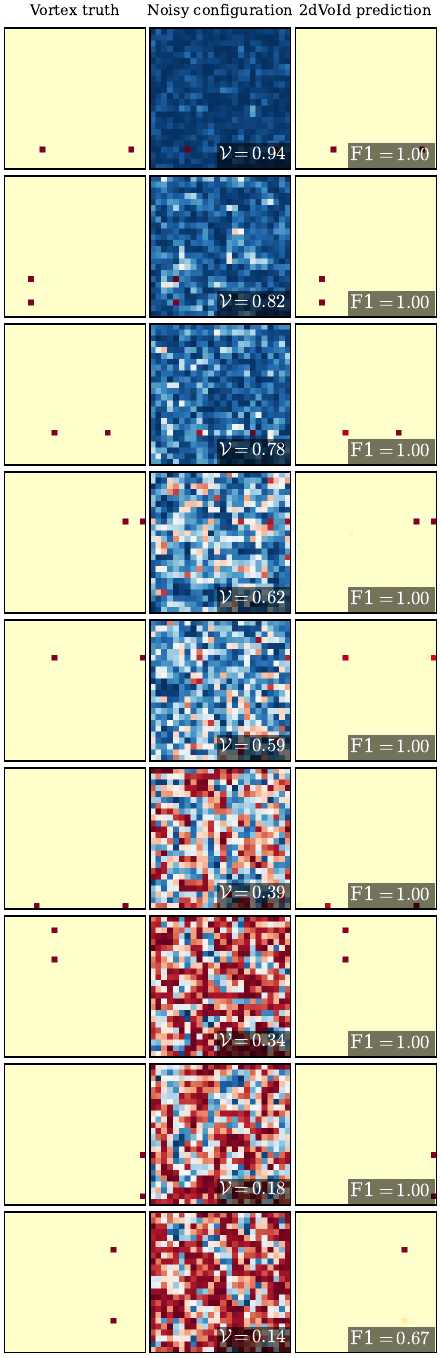} &
\includegraphics[height=0.7\textheight,width=0.32\textwidth]{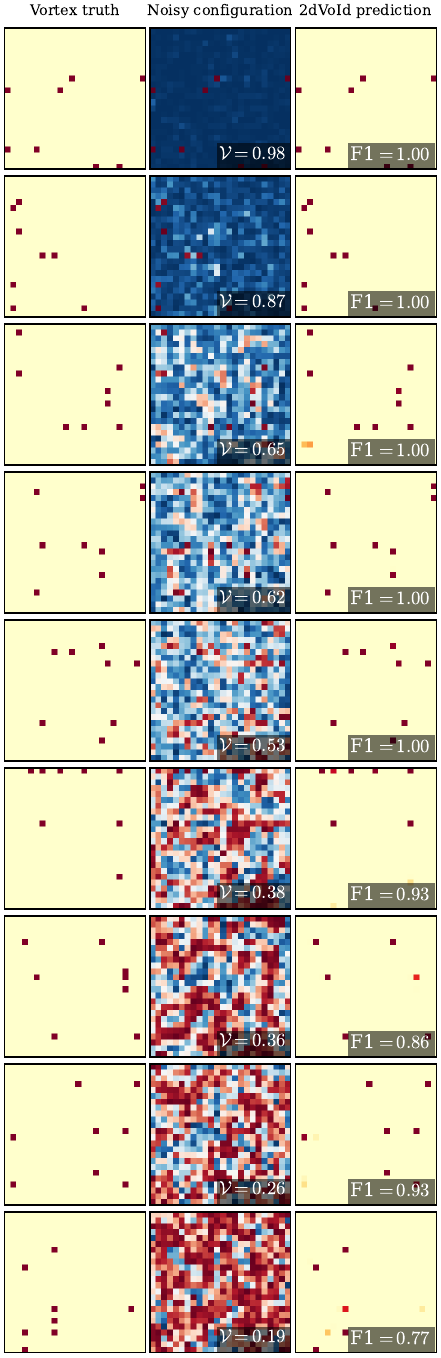} &
\includegraphics[height=0.7\textheight,width=0.32\textwidth]{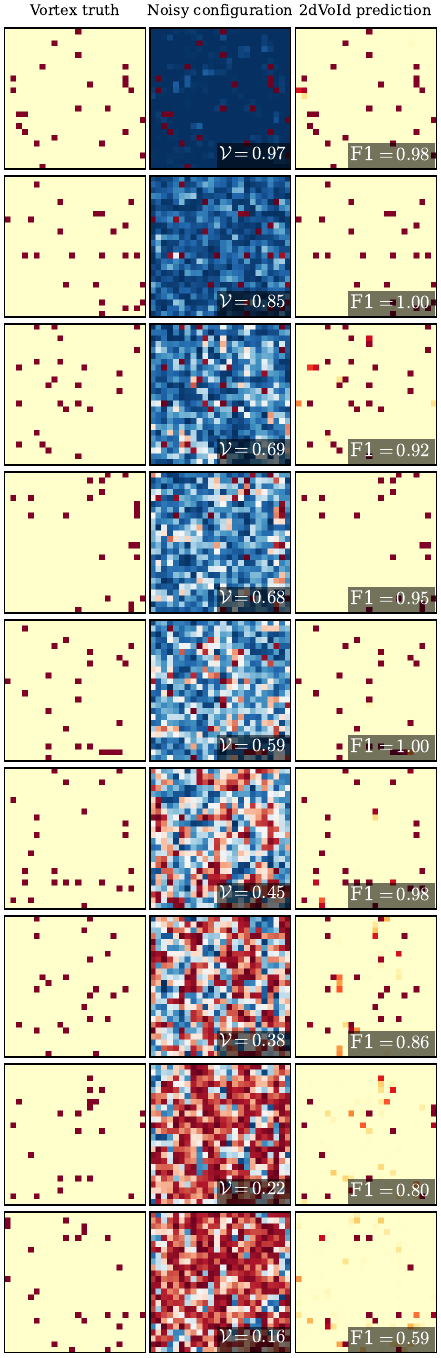}
\end{tabular}
\caption{Gallery of vortex detection examples spanning visibility levels from $\V \approx 0.9$ (top) to $\V \approx 0.1$ (bottom). Each row corresponds to a target visibility level. The figure comprises three panels of three columns each. Left panel: $N_{\rm flip} = 1$, \ie 2 vortices. Center panel: $N_{\rm flip} = 4$, \ie 8 vortices. Right panel: $N_{\rm flip} = 12$, \ie 24 vortices. The three columns in each panel show: (left) the ground-truth vortex mask. Visibility $\V$ and vortex F1 score are annotated on each panel; (center) the cooled plaquette trace after gauge transformations, link noise, and smoothing; (right) the \TwoDVoId probability map, where the color scale is the same as in~\cref{fig:showcase}.}
\label{fig:gallery}
\end{figure*}

\Cref{fig:gallery} shows representative examples of vortex detection by \TwoDVoId on $24 \times 24$ lattices across the full range of visibilities, for 2, 8, and 24 vortices respectively.

Taken together, these results establish that \TwoDVoId reliably identifies center vortices whenever the visibility is moderate or better, across the full range of vortex densities tested. No gauge fixing or center projection is involved at any stage. The vortex visibility $\V$, defined via the effect-size separation of plaquette traces at vortex and non-vortex sites in \cref{eq:visibility}, provides a single-parameter characterization of difficulty that cleanly separates easy from hard configurations regardless of the particular noise and cooling parameters that produced them.

\begin{figure*}[t]
\centering
\includegraphics[width=\textwidth]{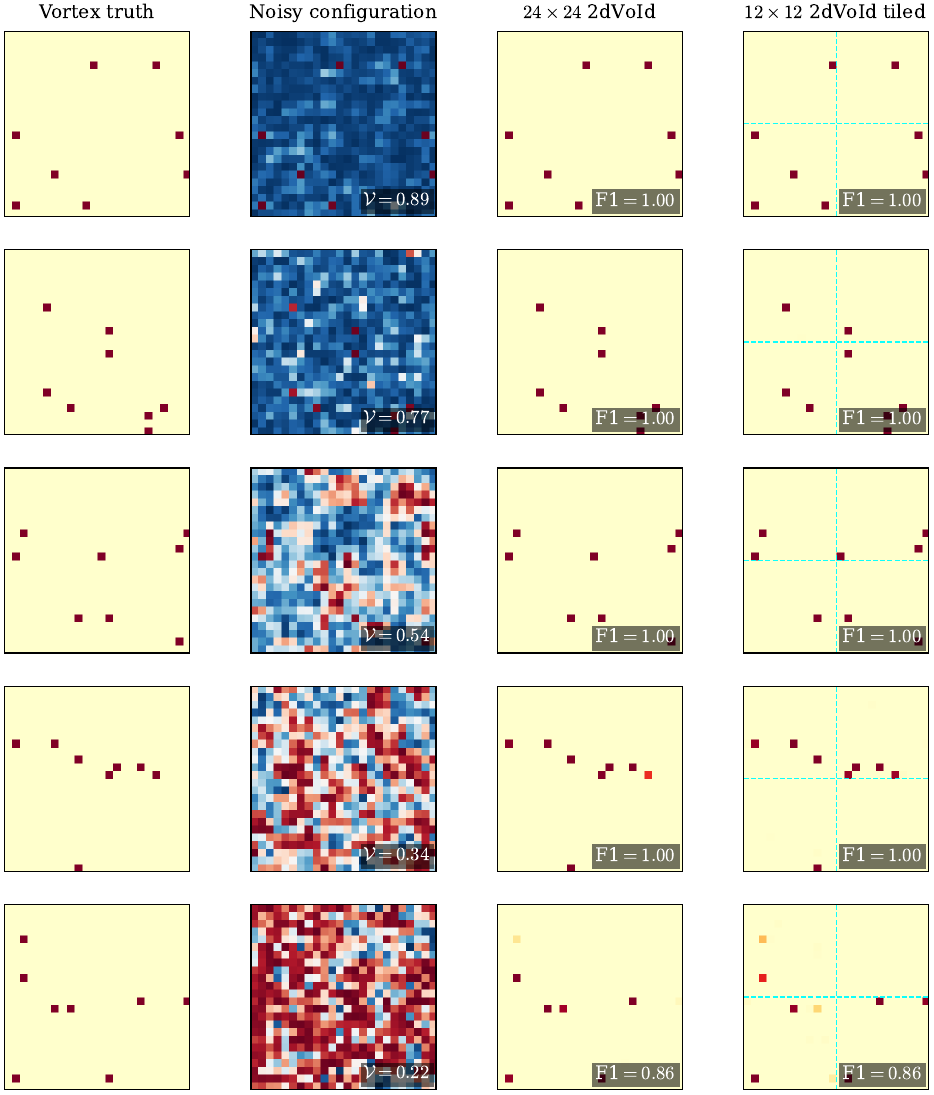}
\caption{Tiling comparison across five visibility levels, with 8 vortices on $24 \times 24$ lattices. The four columns show from left to right: the ground-truth vortex mask; the cooled plaquette trace after gauge transformations, link noise, and smoothing; the full \TwoDVoId probability map; the tiled \TwoDVoId probability map. Color scale is the same as in~\cref{fig:showcase}. The cyan dashed lines on the tiled panel mark quadrant boundaries.}
\label{fig:tiling}
\end{figure*}

We now turn to whether this performance transfers to a model trained on smaller patches, which is a necessary step toward the larger lattice volumes required for realistic simulations in $D=3$ and $D=4$. \Cref{fig:tiling} shows a direct comparison between the native $24 \times 24$ model and the tiled $12 \times 12$ model on the same configurations with 8 vortices, across five visibility levels spanning $\V \approx 0.22$--$0.83$. At high and moderate visibility the two predictions are visually indistinguishable and both match the ground truth. We note that as visibility decreases, the tiled model can begin to fall behind the native model, becoming less confident in the locations of vortices, but the tiled version still performs well. Errors and misidentifications of vortices appear more frequently at the quadrant seams, marked by cyan dashed lines.

\section{Outlook}\label{sec:conclusions}
Our work in this article was restricted to $D=2$ dimensions.  The next step is to generalize the machine learning procedure developed here to higher dimensions, where a pure gauge action can itself generate center vortices.  After training the network on inserted vortices at known locations, the goal is to apply it to configurations generated by lattice Monte Carlo simulations of \SUtwo pure gauge theory and to determine what vortices the network finds.  Given the located vortices, string tensions may be computed by averaging over many configurations. The decisive test will be whether the string tension extracted from the network-identified vortices, averaged over many configurations, agrees with the full string tension of the theory.  This is the central check that has been applied to vortices found by center projection in the maximal center gauge for nearly three decades~\cite{DelDebbio:1996lih,DelDebbio:1998luz}, and the same criterion applies here.  A complementary test is the vortex-removal experiment: configurations from which the network-identified vortices have been removed should exhibit no asymptotic string tension~\cite{deForcrand:1999our}.  Together these provide a stringent operational criterion for whether the network has found the physically relevant excitations. If the procedure in higher dimensions passes these tests, several questions become accessible that have been difficult to settle with center projection alone. The persistent concern with Gribov copies in maximal-center-gauge studies is replaced by sensitivity to the training set and network architecture---a different and arguably better-controlled source of systematics~\cite{Bornyakov:2000ig,Greensite:2004ur}.  Where the network and center projection disagree about which plaquettes carry the confining flux, the comparison is itself informative.  The picture's predictions for the $N$-ality dependence of asymptotic string tensions, for Casimir scaling at intermediate distances~\cite{Faber:1997rp}, and for vortex percolation across the deconfinement transition~\cite{Engelhardt:1999fd} can all be tested against the network's output. Beyond \SUtwo, the obvious extensions are to \SUthree, where the center is $\mathbb{Z}_3$ and the vortex flux is no longer a sign, and to theories with dynamical fermions, where the vortex picture is also expected to organize chiral symmetry breaking~\cite{Trewartha:2015nna,Leinweber:2022ukj}.  The continuum limit raises its own questions: the vortex density located by the network must scale to a finite value in physical units~\cite{Gubarev:2002ek}, and the network's performance must remain stable as the lattice spacing is reduced. 

Each of these directions inherits its difficulty from the underlying physics as well as from the machine learning problem. We also note that in this work we have not made serious attempts to optimize the \TwoDVoId architecture, in the tiled or untiled cases, as such attempts would be premature. Optimization of the number of hidden layers, patch sizes, activation functions, and other hyperparameters should be performed with the full analysis pipeline in mind, and may vary significantly when we move to higher dimensions or a different gauge group. However, we expect that the methodology developed here sets us in the right direction to address those challenges.

\begin{acknowledgments}
This work was supported by the U.S. Department of Energy contract \mbox{DE-AC05-06OR23177}, under which Jefferson Science Associates, LLC operates Jefferson Lab, by U.S.\ Department of Energy Grant Nos.~\mbox{DE-FG02-87ER40365} and~\mbox{DE-SC0013682}, and it contributes to the aims of the U.S. Department of Energy \mbox{ExoHad} Topical Collaboration, contract \mbox{DE-SC0023598}.
\end{acknowledgments}

\bibliographystyle{apsrev4-1}
\bibliography{vortex}   
\end{document}